\documentclass[11pt, a4paper]{article} 

\usepackage[ansinew]{inputenc}
\usepackage{amsmath, amssymb, graphics, amsthm}
\usepackage{epsfig}
\usepackage{color, xcolor}
\usepackage{fancyhdr} 
\usepackage{manfnt}
\usepackage[T1]{fontenc}

\usepackage[normalem]{ulem}

\newcommand{\D}{\underset{\leftarrow}}

\newcommand{\ket}[1]{\left| #1 \right\rangle}

\makeatletter
\@addtoreset{equation}{section}
\makeatother

\newcommand{\be}{\begin{equation}}
\newcommand{\ee}{\end{equation}}
\newcommand{\ba}{\begin{eqnarray}}
\newcommand{\ea}{\end{eqnarray}}

\def\pb#1{\rlap{\lower1.5ex\hbox{$\longleftarrow$}}{#1}}
\def\dpb#1{\rlap{\lower1.5ex\hbox{$\Longleftarrow$}}{#1}}
\def\spb#1{\rlap{\lower1.0ex\hbox{$\leftarrow$}}{#1}}
\def\sdpb#1{\rlap{\lower1.0ex\hbox{$\Leftarrow$}}{#1}}

\title{{\sf Some notes on the Kodama state, maximal symmetry, and the isolated horizon boundary condition}}
\author{
{\sf N. Bodendorfer}\thanks{{\sf 
norbert.bodendorfer@fuw.edu.pl}}\\
\\
{\sf  Faculty of Physics, University of Warsaw, Pasteura 5, 02-093, Warsaw, Poland}\\
}
\date{{\small\sf \today}}

\begin{document} 

\maketitle

{\sf

\begin{abstract}

We recall some well and some less known results about the Kodama state and the related $\theta$ ambiguity in defining canonical variables. Based on them, we make some comments highlighting that the Kodama state for real connection variables can be given a precise meaning and that it implements a vacuum peaked on a (in a suitable sense) maximally symmetric geometry. We also highlight the similarity of this construction with the isolated horizon boundary condition $F \propto \Sigma$ and stress that it is, in agreement with earlier work, inadequate to define the notion of a quantum horizon.

\end{abstract}

}

\section{Introduction}

Loop quantum gravity \cite{RovelliQuantumGravity, ThiemannModernCanonicalQuantum, GambiniAFirstCourse, RovelliBook2} provides a candidate theory for quantum gravity, focussing on the fundamental quantum structure of spacetime. While it offers an intriguing picture of quantum geometry at the smallest scales, it has proven very difficult so far to extract low energy physics from full loop quantum gravity. The main challenge is to understand how classical curved space emerges from a coarse graining of the fundamental dynamics, a formidable task that can be compared to extracting the behaviour of solids from a model of atoms. 

Given the complexity of this problem, even formal proposals for wave functions corresponding to certain classical spacetimes are highly welcome. One of them has been the so called Kodama state \cite{KodamaSpecializationOfAshtekars, SmolinQuantumGravityWith} in the context of complex Ashtekar variables. It is nothing else than the exponential of the Chern-Simons action acting on the vacuum and has been known before in the context of QCD \cite{JackiwTopologicalInvestigationsIn}. In this paper, we will recall some previous results from the literature which show that the Kodama state can be given a precise meaning in the context of real Ashtekar-Barbero variables modified by an additional 1-parameter family of canonical transformations. We will highlight the physical content of this state and point out that it should be appreciated much more, as it provides a simple way of constructing a vacuum which is peaked on a, in a certain sense, maximally symmetric spatial geometry, as opposed to a vanishing spatial metric as in the standard case of Ashtekar-Barbero variables. 

Next, we highlight the similarity of the construction leading to the Kodama state to the isolated horizon boundary condition $F \propto \Sigma$ which famously enters many derivations of black hole entropy in loop quantum gravity. We in particular point out that one can construct a (kinematical) vacuum state on which this condition is implemented on every surface. Comparing to the maximal symmetry enforced in the Kodama state construction, we highlight that the isolated horizon boundary condition does not locate horizons, but merely enforces a partial notion of maximal symmetry. The seminal work \cite{HusainApparentHorizonsBlack} already commented on the inadequacy of $F \propto \Sigma$ for defining quantum horizons, however for different reasons. 

This paper is organised as follows:\\
We start by reviewing some background material on the Kodama state in section \ref{sec:Background}. 
Next, we point out the notion of maximal symmetry enforced by the real Kodama state (section \ref{sec:Com1}), comment on possible solutions to the Hamiltonian constraint (section \ref{sec:Com3}), and compare to the isolated horizon boundary condition (section \ref{sec:Com2}).
In an appendix, we generalise our comments to the higher dimensional connection variables introduced in \cite{BTTVIII}.

\section{Known results on the $\theta$-ambiguity and the Kodama state} \label{sec:Background}

The Ashtekar-Barbero variables \cite{AshtekarNewVariablesFor, BarberoRealAshtekarVariables} are a 1-parameter family of variables which coordinatise the phase space of $3+1$-dimensional general relativity. They are given by a densitised triad $E^a_i$ and an SU$(2)$-connection $A_a^i$, related to geometric variables as 
\be
	q q^{ab} = \beta^2 E^{ai} E^{bj} \delta_{ij}, ~~~ A_a^i = \Gamma_a^i + \beta K_{a}^i, 
\ee
where $q_{ab}$ is the spatial metric, $K_{ab} = K_a^i e_{bi}$ the extrinsic curvature, and $\Gamma_a^i = - \frac 12 \epsilon^{ijk} \Gamma_{jk}$ the spin connection constructed from the undensitized co-triad $e_a^i$ derived from $E^a_i$. $\beta \in \mathbb R \backslash \{0\}$ is a free parameter known as the Barbero-Immirzi parameter \cite{BarberoRealAshtekarVariables, ImmirziQuantumGravityAnd}. 

For the purpose of quantisation using loop quantum gravity techniques, see e.g. \cite{ThiemannModernCanonicalQuantum}, it is of paramount importance that the only non-vanishing Poisson bracket is given by 
\be
	\left\{A_a^i(x), E^b_j(y) \right\} = \delta^{(3)}(x,y) \, \delta_a^b \delta_j^i. \label{eq:PBAE}
\ee
Given this, one can show that the Ashtekar-Lewandowski measure \cite{AshtekarProjectiveTechniquesAnd, AshtekarRepresentationTheoryOf} induces a positive linear functional on the holonomy-flux algebra constructed from $A_a^i$ and $E^a_i$. A Hilbert space representation then follows via the GNS construction. 

On the other hand, one is free to modify the canonical variables as long as \eqref{eq:PBAE} remains the only non-vanishing Poisson bracket.
In particular, a second free parameter $\theta \in \mathbb R$ can be introduced as\footnote{The first reference known to the author where this was explicitly stated in the context of real connection variables is \cite{RezendeThetaParameterIn}, although the idea is immediate given \cite{AshtekarTheCPProblem} and \cite{BarberoRealAshtekarVariables}.}
\be
	P^a_i := E^a_i + \theta \epsilon^{abc} F_{bc}^i, \label{eq:TransTheta} 
\ee
where $F_{bc}^i = - \frac 12 \epsilon^{ijk} F_{bc jk}$ is the curvature of $A_a^i$. It can be checked by direct calculation that
\be
	\left\{A_a^i(x), P^b_j(y) \right\} = \delta^{(3)}(x,y) \, \delta_a^b \delta_j^i \label{eq:PBAP}
\ee
is the only non-vanishing Poisson bracket\footnote{The Poisson bracket of two $P$s integrated over the spatial slice $\Sigma$ against arbitrary smearing functions $\mu_a^i$, $\nu_b^j$ actually evaluates to $\{P^a_i[\mu_a^i],P^b_j[\nu_b^j]\} = \theta \int_{\partial \Sigma} \mu_a^i \nu_{bi} \epsilon^{ab} d^2x$. In order for it to vanish, we need to choose the smearing functions of the $P$s to vanish on $\partial \Sigma$, or simply work on a spatial slice without boundary.}. One thus has a $2$-parameter family of connection variables labelled by $\beta$ and $\theta$.

The idea for \eqref{eq:TransTheta} dates back to Yang-Mills theory and its $\theta$-ambiguity \cite{JackiwTopologicalInvestigationsIn}. Within the loop quantum gravity literature, it was first\footnote{While \cite{KodamaSpecializationOfAshtekars} was printed before \cite{AshtekarTheCPProblem}, the available preprint of \cite{AshtekarTheCPProblem} predates the submission date of \cite{KodamaSpecializationOfAshtekars}.} discussed in \cite{AshtekarTheCPProblem} and \cite{KodamaSpecializationOfAshtekars, KodamaHolomorphicWaveFunction}, see also \cite{SmolinTheChernSimonsInvariant, SmolinQuantumGravityWith, SooFurtherSimplificationOf, RandonoAGeneralizationOf, RandonoGeneralizingTheKodamaI, RandonoGeneralizingTheKodamaII} for more recent work. Mostly, it is approached from the point of view of the so called Kodama state 
\be
	\ket{\Lambda} := \ket{\exp \left( \frac{3}{2\Lambda} \int_\Sigma  S_{\text{CS}}(A) \right)} \label{eq:KodamaState},
\ee
which has originally been a proposal for a physical wave function within the context of self-dual Ashtekar variables. This state is nothing else than the exponential of the Chern-Simons functional $S_{\text{CS}} = \text{Tr} \left( A \wedge d A + \frac 23 A \wedge A \wedge A\right)$ integrated over the spatial slice $\Sigma$. 
Formally, the Hamiltonian constraint with a cosmological constant in self-dual variables annihilates this state, since
\be
	\hat H \ket{\Lambda} =  \epsilon^{ade} \frac{\delta}{\delta A_d^l}   \frac{\delta}{\delta A_e^m} \epsilon^{ilm} \left(\frac {2\Lambda}{3}  \frac{\delta}{\delta A_a^i}+  \epsilon^{abc} \hat F_{bc}^i \right) \ket{\Lambda}=0.
\ee
There are however several technical problems which have prohibited to make this precise, including that there is so far no Hilbert space representation of the self-dual variables which implements their reality conditions, see \cite{ThiemannModernCanonicalQuantum} for more discussion. Moreover, the direct analogue of the Kodama state for complex Ashtekar variables in Yang-Mills theory has unphysical properties \cite{WittenANoteOn}. 

On the other hand, the Kodama state has much better properties once it is considered in the context of the real Ashtekar-Barbero variables \cite{RandonoAGeneralizationOf, RandonoGeneralizingTheKodamaI, RandonoGeneralizingTheKodamaII}. In fact, adding an additional $i$ in the exponent of \eqref{eq:KodamaState}, the state becomes oscillatory and formally generates the canonical transformation \eqref{eq:TransTheta} as\footnote{Note that in the case of real variables, $\hat E^a_i = - i \frac{\delta}{\delta A_a^i}$, whereas in self-dual variables, $\hat E^a_i =   \frac{\delta}{\delta A_a^i}$.}
\be
	\hat P^{a}_i = \exp \left( - i \theta \int_\Sigma  S_{\text{CS}}(\hat A) \right)~ \hat E^a_i ~ \exp \left( i \theta \int_\Sigma  S_{\text{CS}}(\hat A) \right) \text{.}
\ee
There thus exists a rigorous way to define the {\it real} Kodama state as follows \cite{RezendeThetaParameterIn}. Instead of quantising using holonomies and fluxes constructed form $A_a^i$ and $E^a_i$, we construct fluxes from $P^a_i$ instead of $E^a_i$. On the resulting holonomy-flux algebra, we define the standard Ashtekar-Lewandowski state. The fact that fluxes constructed from $P^a_i$ annihilate the vacuum then directly corresponds to the relation $E^a_i + \theta \epsilon^{abc} F_{bc}^i=0$, which one would formally expect from acting on the real variable version of \eqref{eq:KodamaState} in the standard quantisation based on $A_a^i$ and $E^a_i$. From the $A_a^i$, $E^a_i$ perspective, this means that we have defined the Kodama state \eqref{eq:KodamaState} to have unit norm and that only the combination $E^a_i + \theta \epsilon^{abc} F_{bc}^i$ smeared over a two-surface annihilates it, while holonomies act via multiplication as usual. In particular, this means that the usual flux derived from $E^a_i$ does not have a  well defined action on the Kodama state\footnote{It is possible however to regularise the field strength in terms of holonomies \cite{SahlmannBlackHoleHorizons}.}.

\section{Comments} 

\subsection{A maximally symmetric vacuum} \label{sec:Com1}

The main virtue of the Kodama state advocated in \cite{SmolinQuantumGravityWith} was that it (formally) constitutes a non-trivial ground state of the theory corresponding to de Sitter space. Due to the mathematical problems associated with complex Ashtekar variables, this proposal did not receive much attention. As we have seen in the previous section, a rigorous implementation of the Kodama state can however be achieved using the classical canonical transformation \eqref{eq:TransTheta}. While the real form of the Kodama state was originally introduced in Randono's work \cite{RandonoAGeneralizationOf, RandonoGeneralizingTheKodamaI, RandonoGeneralizingTheKodamaII}, we know of \eqref{eq:TransTheta} as first being spelled out in \cite{RezendeThetaParameterIn} in the context of real variables. The main point of \cite{RezendeThetaParameterIn} was however to show that using the isolated horizon boundary condition, one can define the area operator derived from the geometric flux $E^a_i$ also in the context of $A_a^i$, $P^a_i$ variables and apply this to the black hole entropy computation\footnote{A different interpretation of this computation has been proposed in \cite{BNII}, where it is explained that the appropriate Wald entropy is obtained as a result of using fluxes constructed from $P^a_i$.}.  
Here, we want to stress a different aspect of the choice of variables \eqref{eq:TransTheta}, which is very close to Smolin's proposal \cite{SmolinQuantumGravityWith}. 

As a short digression, let us mention again that the problem of obtaining the classical limit has proven very hard in loop quantum gravity. In particular, this is connected to the problem of finding a quantum state which describes some classical spacetime, say Minkowski space. Such a state can in principle be an arbitrary superposition of spin network states, and it is not clear how to construct it or how to properly extract the low energy physics from it. Starting from the Asthekar-Lewandowski vacuum in terms of $A_a^i$, $E^a_i$ variables, this is in particular very complicated because one needs to build up space completely from spin networks, since the fluxes identically vanish on the vacuum state, see e.g. \cite{AshtekarWeavingAClassical, BombelliStatisticalGeometryOf}. One can improve on this situation for example by considering background fluxes as in \cite{KoslowskiDynamicalQuantumGeometry, KoslowskiLoopQuantumGravity}, or change the vacuum to one peaked on flat connections \cite{DittrichANewVacuum, BahrANewRealization} and try to approach the problem from there. Another interesting recent approach consists in defining condensate states \cite{GielenCosmologyFromGroup}. 

Let us now turn to the quantum theory obtained by standard means from the $A_a^i$, $P^a_i$ variables. Since the Ashtekar-Lewandowski vacuum is annihilated by all fluxes, it implements the relation 
\be
	P^a_i = E^a_i + \theta \epsilon^{abc} F_{bc}^i = 0 \label{eq:VacuumCondition}
\ee
While for $\theta = 0$ this would correspond to a vanishing spatial metric, we can flesh out the SU$(2)$ invariant content of \eqref{eq:VacuumCondition} as 
\begin{eqnarray}
	q_{c[a} q_{b]d} = \beta \theta F_{ab ij} e^c_i e^d_j := \beta \theta  F_{abcd} &=& \beta \theta  R^{(3)}_{ab cd}  - \theta \epsilon^{ijk} e_{cj} e_{dk} \left( 2 \beta^2 \nabla_{[a} K_{b]}^i + \beta^3 \epsilon^{imn} K_{am} K_{bn} \right) \nonumber \\
		&=&  \beta \theta  R^{(3)}_{ab cd}  - 2 \beta^2 \theta \epsilon_{cde}  \sqrt{q} \nabla_{[a} K_{b]} {}^e - 2 \beta^3 \theta K_{c[a} K_{b]d}   \label{eq:SymmetryCondition}
\end{eqnarray}
by contracting it with a triad and performing some simple rewriting, where $\nabla_a$ acts on internal indices with the spin connection $\Gamma_{ai}(e)$ and on tensor indices with the spatial Christoffel symbols. We see that \eqref{eq:VacuumCondition} imposes that the spatial slice is maximally symmetric in the sense of the Ashtekar-Barbero connection $A_a^i$ satisfying \eqref{eq:SymmetryCondition} and note that this condition has an explicit dependence on the Barbero-Immirzi parameter $\beta$ which cannot be absorbed in $\theta$.
If in addition the extrinsic curvature would vanish, then the spatial slice would indeed be maximally symmetric in the usual sense, because then $q_{c[a} q_{b]d} \propto R^{(3)}_{ab cd}$. The Ashtekar-Lewandowski vacuum based on $A_a^i$, $P^a_i$ variables thus seems to be very useful if one is interested in spacetimes obeying symmetry conditions such as \eqref{eq:VacuumCondition}.

A straight forward example would be to consider the spatial slice to be a three-sphere with standard (maximally symmetric) metric and vanishing extrinsic curvature. In order to satisfy the Hamiltonian constraint, we would need to introduce a positive cosmological constant. This slice is then nothing but the $t=0$ slice of de Sitter space in closed slicing, similar to the arguments in \cite{SmolinQuantumGravityWith}, where a flat slicing was employed. 
The vector constraint is automatically solved due to the vanishing extrinsic curvature.

As noted in \cite{RezendeThetaParameterIn}, in order to construct the Hamiltonian constraint, one needs to improve on the methods developed in the context of $A_a^i$, $E^a_i$ variables, since the usual geometric operators corresponding to area and volume are not well defined any more. However, one can now build similar operators by substituting $E^a_i$ with $P^a_i$ and find a similar regularisation of the Hamiltonian constraint. The problem is essentially already solved once an operator corresponding to the physical volume has been constructed which vanishes on degenerate vertices, since then one can simply repeat Thiemann's construction using this operator. We will briefly outline one possibility how this can be achieved. Others, in particular simpler ones, might exist.   

First, we construct the volume operator $V_p$ build from $P^a_i$ by following \cite{AshtekarDifferentialGeometryOn}. It has the same properties as the standard volume operator. We first define $p^2 := \frac 16 \epsilon^{abc} \epsilon_{ijk} P^a_i P^b_j P^c_k$ and $V_p(R) = \int_R \sqrt{p^2}$. Now, following \cite{ThiemannQSD1}, we can make use of the Poisson bracket identity 
\be
	\left\{A_a^i, V_p(R) \right\} := \frac 12 p_a^i = \frac{1}{2 V_p(R)} \left( \sqrt{q} e_a^i - 2 \theta F^j_{ab} P^b_k \epsilon^{ijk} + \theta^2 F_{ab}^j F_{cd}^k \epsilon^{ijk} \epsilon^{bcd} \right) \label{eq:pai}
\ee
where $R$ contains the point at which $A_a^i$ is evaluated.
First, we can now build an operator corresponding to the inverse $p$-volume by using \eqref{eq:pai} in $p = |\frac 16 \epsilon^{abc} \epsilon_{ijk} p_a^i p_b^j p_c^k|$ and taking suitable roots of the $p$-volumes. Next, we can build an operator corresponding to $e_a^i \sqrt{q}/\sqrt{p}$ by subtracting from $p_a^i$ the terms proportional to field strengths, ordering the $1/V$ terms to the right so that they annihilate degenerate vertices. Proceeding similarly, we can build an operator corresponding to $V_q^4 / V_p^3$, where $V_q$ is the desired volume obtained from integrating $\sqrt{q} = |\frac 16 \epsilon^{abc} \epsilon_{ijk} e_a^i e_b^j e_c^k|$. Multiplication by $V_p^3$ and taking of the fourth root then gives $V_q$. 
A technicality which we have so far not looked at is that $V_q^4$ has to be self-adjoint, so that we can take the square root. Also, one would have to check whether anomaly freedom in the sense of \cite{ThiemannQSD1} still holds. We will leave this for future research, as we already see a strong enough motivation further study.

\subsection{Solving the Hamiltonian constraint?} \label{sec:Com3}

We briefly note that one can in principle construct a regularisation of the Hamiltonian constraint with a cosmological constant which annihilates the Ashtekar-Lewandowski vacuum build from $A_a^i$, $P^a_i$ variables, following Smolin's original proposal \cite{SmolinQuantumGravityWith} for complex Ashtekar variables. Given a cosmological constant, one can set $\theta$ such that the Hamiltonian constraint schematically reduces to
\be
	H[N] = E E (E+\theta F) + (1+\beta^2) K^2 \text{.} \label{eq:HamConstr}
\ee
In the quantisation, one would now simply order $P^a_i := E^a_i + \theta \epsilon^{abc} F_{bc}^i$ to the right in the first term, so that it annihilates the vacuum. The second term also automatically vanishes on the vacuum due to the standard regularisation procedure \cite{ThiemannQSD1}. 

Should we fully trust this construction? The answer to this question seems in the negative, since the vanishing of the extrinsic curvature terms stems from the specific regularisation chosen. It would be more satisfactory to have both terms in \eqref{eq:HamConstr} cancel each other, as opposed to vanishing individually, in particular since they should not commute. It thus seems too early to consider this implementation of the real Kodama state as a satisfactory solution to all the quantum constraints. However, it is certainly very interesting from the point of view of providing a vacuum state corresponding to a non-degenerate geometry. 

We note that in the context of Euclidean gravity, the second term in \eqref{eq:HamConstr} would be absent if we would choose $\beta = 1$, and thus the above problem would be avoided.

\subsection{The isolated horizon boundary condition} \label{sec:Com2}

Equation \eqref{eq:TransTheta} is structurally very similar to the isolated horizon boundary condition \cite{SmolinLinkingTopologicalQuantum, AshtekarIsolatedHorizonsThe, EngleBlackHoleEntropyFrom}
\be
	{F}_{\D{bc}}^i = - c \beta E^{a i} \epsilon_{a\D{bc}}, \label{eq:IHBC}
\ee
used in the context of black hole in loop quantum gravity\footnote{In computations of black hole entropy following \cite{AshtekarQuantumGeometryOf}, one is actually not imposing \eqref{eq:IHBC}, but only that the curvature in \eqref{eq:IHBC} derives from the connection that one uses in the boundary symplectic structure \cite{BII}. This is in line with the entanglement entropy interpretation of this computation \cite{HusainApparentHorizonsBlack, DonnellyEntanglementEntropyIn, BII}. A computation directly implementing \eqref{eq:IHBC} is given in \cite{SahlmannBlackHoleHorizons}, see also \cite{OritiHorizonEntropyFrom}.}. In fact, the condition $P^{ai} = 0$ enforced by the vacuum implements \eqref{eq:IHBC} on any two-surface after a suitable identification of the parameters $c$ and $\theta$. However, as seen before, the physical content of $P^{ai}=0$ is to enforce a notion of maximal symmetry in the sense of \eqref{eq:SymmetryCondition}, which is independent of the notion of a horizon. 
This analysis thus strengthens the result of \cite{HusainApparentHorizonsBlack} that a quantised version of \eqref{eq:IHBC} should not be used to define a quantum horizon.

Another instructive counterexample to \eqref{eq:IHBC} selecting a horizon is provided by considering the Schwarzschild black hole of mass $m$ in standard $(t,r,\theta,\phi)$ coordinates, where the extrinsic curvature vanishes. Here, on spheres of constant $r$, equation \eqref{eq:IHBC} reduces to 
\be
	q_{\theta [\theta} q_{\phi] \phi} = \frac 12 r^4 \sin^2 \theta = \frac{\beta}{2c} R_{\theta \phi \theta \phi} = \frac{ \beta m r \sin^2 \theta}{c} \text{.}
\ee
This means that at the sphere $r = \sqrt[3]{\frac{2 \beta m}{c}}$, \eqref{eq:IHBC} is satisfied. 
For fixed c, we can thus choose the ratio $2m/r$ arbitrary, while still satisfying \eqref{eq:IHBC}. Yet another example is given by stationary cylinders of constant radius embedded in flat spacetime\footnote{We thank an anonymous referee for pointing out this example.}.

\section{Conclusion}

In this paper, we have recalled some known facts about the Kodama state and especially its implementation in the context of real variables. We pointed that it has the very useful property of providing us with a vacuum describing a highly symmetric and non-degenerate geometry. An at first seemingly unrelated research topic within loop quantum gravity is the computation of black hole entropy, and in this context the implementation of the isolated horizon boundary condition $F \propto \Sigma$.  We emphasised the strong similarity of this boundary condition with the implementation of the real Kodama state using a classical canonical transformation. In particular, it was highlighted that the isolated horizon boundary condition $F \propto \Sigma$ is imposing a part of a maximal symmetry condition as opposed to characterising horizons. Its quantisation therefore does not serve as an appropriate definition of a quantum horizon.

\section*{Acknowledgements}
This work was supported by the Polish National Science Centre grant No. 2012/05/E/ST2/03308. Discussions with Hanno Sahlmann and Thomas Zilker are gratefully acknowledged.

\begin{appendix}
\section{Higher dimensions}

Next to the Ashtekar-Barbero variables, there exists another set of canonical variables, in terms of which loop quantum gravity can be constructed in $D+1$ dimensions for $D \geq 2$ \cite{BTTI, BTTII, BTTIII, BTTIV}. The variables are an SO$(D+1)$ connection $A_{aIJ}$ with conjugate momentum $\pi^{aIJ}$, where $I,J = 0, \ldots D$. They are related to geometric variables as $2 q q^{ab} = \beta^2  \pi^{aIJ} \pi^{bKL} \delta_{IK} \delta_{JL}$ and $A_{aIJ} = \Gamma_{aIJ} + 2 \beta n_{[I} K_{a|J]} + \text{gauge}$, where $n^I$ is a normal constructed from $\pi^{aIJ}$ and $\Gamma_{aIJ}$ is the Peldan hybrid spin connection \cite{PeldanActionsForGravity}. In addition to the Hamiltonian, spatial diffeomorphism, and Gau{\ss} constraint, there is an additional simplicity constraint $\pi^{a[IJ} \pi^{b|KL]}=0$, which enforces that $\pi^{aIJ} = 2/\beta \, n^{[I} E^{a|J]}$ with $n_I E^{aI}=0$, i.e. $\pi^{aIJ}$ derives from a $(D+1)$-bein related to the D-dimensional spatial metric as $q q^{ab} = E^{aI} E^{bJ} \delta_{IJ}$.

We will now generalise the discussion of this paper to odd $D$. The analogue of the canonical transformation \eqref{eq:TransTheta} reads
\be
	P^{aIJ} = \pi^{aIJ} + \theta \epsilon^{a b_1 c_1 \ldots b_n c_n} \epsilon^{IJ K_1 L_1 \ldots K_n L_n} F_{b_1 c_1 K_1 L_1} \ldots F_{b_n c_n K_n L_n} \label{eq:HDTrans}
\ee
with $n = (D-1)/2$. In the computation, some additionally appearing terms as opposed to \eqref{eq:TransTheta} vanish by the Bianchi identity. The condition $P^{aIJ}=0$ translates to 
\be
	\delta_{[a_1}^{b_1} \delta_{c_1}^{d_1} \ldots \delta_{a_n}^{b_n} \delta_{c_n]}^{d_n} = \theta (D-1)! F_{[a_1} {}^{b_1} {}_{c_1} {}^{d_1} \ldots F_{a_n} {}^{b_n} {}_{c_n]} {}^{d_n} \label{eq:HDCond1}
\ee
and
\be
	e^d_{K} e^e_L \epsilon^{c a_1 b_1 \ldots a_n b_n}  \epsilon^{KL I_1 J_1 \ldots I_n J_n} F_{a_1b_1I_1J_1} \ldots F_{a_n b_n I_nJ_n} = 0 \label{eq:HDCond2}
\ee
where \eqref{eq:HDCond1} comes from projecting \eqref{eq:HDTrans} along the $n^I$ direction and \eqref{eq:HDCond2} from the directions orthogonal to $n^I$. Again, \eqref{eq:HDTrans} = 0 on a $D-1$ surface is equivalent to the isolated horizon boundary condition for spherically symmetric and non-distorted isolated horizons derived in \cite{BTTXII}. We can thus again choose to construct the quantum theory based on holonomies and fluxes derived form $A_{aIJ}$, $P^{aIJ}$. The Ashtekar-Lewandowski vacuum in this case would again satisfy the isolated horizon boundary condition for arbitrary $D-1$ surfaces for a suitable choice of $\theta$.

\end{appendix}


\end{document}